# Predicting Two-Dimensional Boron-Carbon Compounds by the Global Optimization Method

Xinyu Luo,[1] Jihui Yang,[1] Hanyu Liu,[2] Xiaojun Wu,[3] Yanchao Wang,[2] Yanming Ma,[2] Su-Huai Wei,[4] Xingao Gong,[1] and Hongjun Xiang[1, *]

[1]Key Laboratory of Computational Physical Sciences (Ministry of Education), and Department of Physics, Fudan University, Shanghai 200433, P. R. China

[2]State Key Lab of Superhard Materials, Jilin University, Changchun 130012, China

[3]Department of Material Science and Engineering, Hefei National Laboratory for Physical Science at the Microscale and CAS Key Lab of Materials for Energy Conversion, University of Science and Technology of China, Hefei, Anhui, 230026, China

[4]National Renewable Energy Laboratory, Golden, Colorado 80401, USA

## Abstract

We adopt a global optimization method to predict two-dimensional (2D) nanostructures through the particle-swarm optimization (PSO) algorithm. By performing PSO simulations, we predict new stable structures of 2D boron-carbon (B-C) compounds for a wide range of boron concentrations. Our calculations show that: (1) All 2D B-C compounds are metallic except for $BC_3$ which is a magic case where the isolation of carbon six-membered ring by boron atoms results in a semiconducting behavior. (2) For C-rich B-C compounds, the most stable 2D structures can be viewed as boron doped graphene structures, where boron atoms typically form 1D zigzag chains except for $BC_3$ in which boron atoms are uniformly distributed. (3) The most stable 2D structure of BC has alternative carbon and boron ribbons with strong in-between B-C bonds, which possesses a high thermal stability above 2000K. (4) For B-rich 2D B-C compounds, there is a novel planar-tetracoordinate carbon motif with an approximate $C_{2v}$ symmetry.

## INTRODUCTION

Graphene, a two-dimensional (2D) single layer of carbon atoms arranged in a honeycomb lattice, has been the focus of recent research efforts,[1-4] due to its unique zero-gap electronic structure and the massless Dirac Fermion behavior. The unusual electronic and structural properties make graphene a promising material for the next generation of faster and smaller electronic devices. The need for miniaturization of electronic devices calls for continued development of new materials with reduced dimensionality. Besides graphene, some other 2D materials were fabricated as well. Recently, Coleman *et al.*[5] reported that a straightforward liquid exfoliation technique can efficiently produce monolayer 2D nanosheets from a variety of inorganic layered materials such as BN, $MoS_2$ and $WS_2$. There is a growing interest in exploring the structures and properties of boron nanostructures because boron possesses a richness of chemistry. This element has been extensively investigated both theoretically[6,7] and experimentally[8] in various forms (e.g., bulk boron, nanotubes,[9] clusters and quasi planar sheets). The general perception of a monolayer boron sheet is that it occurs as a buckled sheet with a triangular lattice.[10] Recently, monolayers of boron comprised of triangular and hexagonal motifs, known as "α-sheets",[11,12] have been predicted to be energetically more stable than the flat triangular sheets, which has not been experimentally synthesized. This leads to the postulation of the existence of the $B_{80}$ fullerenes.[13] The stability and electronic properties of boron nanoribbons were also explored.[14] These novel nanostructures with a number of interesting properties offer numerous potential applications in tribology, high-energy-density batteries, sensors, photoconversion of solar energy and nanoelectronics. In particular, 2D materials are of special importance because they are usually parent structures of one-dimensional nanotubes and zero-dimensional nanocages.

Some 2D boron-carbon (B-C) alloy structures were studied both experimentally and theoretically. A novel material with the composition $BC_3$ with a graphite-like structure was confirmed by electron diffraction data.[15] This $BC_3$ honeycomb sheet with excellent crystalline quality was grown uniformly over a macroscopic surface area of $NbB_2$ (0001).[16] Theoretically, a first-principles study[17] showed that a monolayer of $BC_3$ is an indirect-gap semiconductor. By varying the ratio between benzene and $BCl_3$, $B_xC_{1-x}$ compounds with x = 0.17 have been synthesized at 900℃ by Way *et al.*[18] who suggested the existence of an ordered $BC_5$ compound and proposed a possible structure of $BC_5$. The speculated monolayer $BC_5$ was later predicted to be metallic by density functional calculations.[19] On the boron-rich side, Wu *et al.* recently proposed a 2D $B_2C$ sheet in which the boron and carbon atoms are packed into a mosaic of hexagons and rhombuses.[20] In this 2D $B_2C$ graphene, each carbon atom is bonded with four boron atoms, forming a planar-tetracoordinate carbon (ptC) moiety.[20] Despite the recent discovery of these novel 2D B-C nanostructures, a complete understanding of the structures and properties of 2D B-C compounds with a wide range of boron concentrations is still elusive. This is stemmed from the fact that the 2D boron sheet differs from graphene, making the prediction of the structures of 2D B-C compounds extremely difficult.

In this work, we propose a general global optimization method to predict 2D nanostructures based on the particle swarm optimization (PSO) technique as implemented in the Crystal structure AnaLYsis by Particle Swarm Optimization (CALYPSO) code.[21] Our extensive tests show that the new method is very efficient in finding the stable 2D nanostructures. Utilizing our method, we study 2D $B_xC_y$ compounds with several B concentrations. Our simulations reveal new 2D ground state structures of $BC_5$, $BC_2$, $BC$, $B_2C$, $B_3C$, and $B_5C$. We show that 2D C-rich B-C compounds adopt the graphene-like honeycomb structure, therefore, can be treated as B doped graphene, the B-rich compounds has less similarity with that of boron α-sheet, although they both consist of different arrangement of hexagons and triangles. It is also interesting to see that a common feature of B-rich B-C compounds is that they all have similar $C_{2v}$-like ptC motifs.

## METHODS

### PSO algorithm for 2D system

Previously, several approaches have been proposed to predict the ground state structures of crystals and clusters.[22] Recently, we developed a method for 3D crystal structure prediction through PSO algorithm within the evolutionary scheme.[21] PSO is designed to solve problems related to multiple dimensional optimization,[23] which is inspired by the social behavior of bird flocking or fish schooling. The key idea is to have a swarm of interacting particles, each representing a candidate solution to a given optimization problem. Thus, particles are embedded in the search space and explore the solution space by flying around. Moreover, the particles are also attracted to high fit regions located by other particles. Recently, we have applied PSO algorithm into the field of crystal structure prediction for materials with the only known information of chemical compositions at given external conditions (e.g. pressure) as implemented in CALYPSO code.[21] In this application, each particle is treated as a specific structure in a high dimensional space which adjusts its own flying according to its flying experience as well as the flying experience of other structures. The system is initialized with a population of random structural solutions and searches for optima by updating generations. We have demonstrated that PSO algorithm on crystal structure prediction is highly efficient with a faster and cheaper way compared with other methods.[21] The CALYPSO method has been successful in predicting structures for various high-pressure systems including the semiconducting phase of lithium,[24] the electride structure of Mg,[25] and the superconducting phases of $Bi_2Te_3$.[26]

Our earlier PSO algorithm is specially designed for 3D crystal structure prediction.[21] Here, we have, for the first time, applied the PSO algorithm to 2D systems. In this application, we only consider single atomic layer 2D systems. It should be noted that a straightforward extension of our method can be applied to 2D systems with finite thickness, e.g. $MoS_2$. In our method, we first generate a set of random 2D structures with a randomly chosen symmetry. Different from the 3D crystal case where the 230 space groups are used, we randomly select a 2D symmetry group among the 17 plane space groups. Once a particular plane space group is selected, the lateral lattice parameters are then confined within the chosen symmetry. The atomic coordinates are randomly generated with the imposed symmetry constrain. The generation of random structures ensures unbiased sampling of the energy landscape. The explicit application of symmetric constraints leads to significantly reduced search space and optimization variables, and thus fastens global structural convergence. Subsequently, local optimization including the atomic coordinates and lateral lattice parameters is performed for each of the initial structures. In the next generation, a certain number of new structures (the best 60% of the population size) are generated by PSO. The other structures are generated randomly, which is critical to increase the structure diversity. Within the PSO scheme, a structure in the searching phase space is regarded as a particle. A set of particles (structures) is called a population or a generation. The positions of the particle are updated according to the following equation:

$$x_{i,j}^{t+1} = x_{i,j}^{t} + v_{i,j}^{t+1}$$

where x and v are the position and velocity, respectively (i is the atom index, j refers to the dimension of structure with j∈{1, 2}, and t is the generation index). The new velocity of each particle is calculated based on its previous location $x_{i,j}^{t}$ before optimization, previous velocity $v_{i,j}^{t}$, current location $pbest_{i,j}^{t}$ with an achieved best fitness (i.e., lowest energy), and the population global location $gbest_{i,j}^{t}$ with the best fitness value for the entire population:

$$v_{i,j}^{t+1} = \omega v_{i,j}^{t} + c_1 r_1 (pbest_{i,j}^{t} - x_{i,j}^{t}) + c_2 r_2 (gbest_{i,j}^{t} - x_{i,j}^{t})$$

where ω (in the range of 0.9-0.4) denotes the inertia weight, $c_1$ = 2 and $c_2$ = 2, $r_1$ and $r_2$ are two separately generated random numbers and uniformly distributed in the range [0, 1]. The initial velocity is generated randomly. All the structures produced by the PSO operation are then relaxed to the local minimum. Usually, tens of iterations are simulated to make sure that the lowest energy structures are found. By symmetry, the local optimization and the PSO operation keep the 2D nature of the structures. We have implemented the PSO algorithm for 2D systems into the CALYPSO code.[21]

In our calculations, we usually set the population size to 30. We consider all possible cell sizes with the total number of atoms no more than 20. The number of generations is fixed to 30.

### DFT calculations

In the PSO simulations, we use density functional theory (DFT) to relax the structures and calculate the energies. In the DFT plane-wave calculations, we use the local density approximation (LDA). The ion-electron interaction is treated using the projector augmented wave (PAW)[27] technique as implemented in the Vienna ab initio simulation package (VASP).[28] For relaxed structures, the atomic forces are less than 0.01eV/Å. Because the 3D periodic boundary condition is adopted in VASP, we simulate the 2D systems by constrain all the atoms in an ab-plane which is perpendicular to the c lattice vector with the length fixed to 10Å. For the Brillouin zone integration, we generate the n×m×1 k-mesh according to the Monkhorst-Pack scheme, where n and m are determined by the lateral lattice constant. The phonon calculations are performed using the direct method as implemented in the Phonopy program.[29]

## RESULTS AND DISCUSSION

### Known 2D Systems

We first apply the designed PSO method through CALYPSO code to predict the most stable 2D structure of carbon. As expected, the global optimization method successfully predicts the graphene structure with a two-atom unit cell by only one generation. We also find the most stable 2D hexagonal BN[30] structure within one generation. Although bulk ZnO takes the wurtzite structure, previous experiment showed that single-layer ZnO[31] has a similar planar structure as BN. We perform four separate PSO simulations to find the most stable 2D structure of ZnO. All four simulations predict the correct most stable hexagonal ZnO structure: Three of them use only one generation, while the hexagonal structure emerges in the third generation in the other simulation. For boron, we again find the known most stable 2D structure, i.e., the α-sheet with eight B atoms,[11,12] within only two generations. These benchmarks suggest that our PSO algorithm is rather effective in predicting stable 2D materials.

**2D Boron-Carbon Compounds**

We consider the 2D B-C compounds with seven different B concentrations: $BC_5$, $BC_3$, $BC_2$, $BC$, $B_2C$, $B_3C$, and $B_5C$. The low energy 2D structures for B-C compounds predicted from our PSO simulations via CALYPSO code are shown in Figs. 1-7. We use I, II, III, …, to name the structures in the order of increasing energy. We can see that for carbon-rich compounds ($BC_2$, $BC_3$, and $BC_5$), the 2D sheets can be viewed as boron doped graphene structures. Interestingly, there are isolated 1D zigzag boron chains in the most stable 2D structures of $BC_5$ (i.e., $BC_5$-I) and $BC_2$ (i.e., $BC_2$-I). For $BC_5$, it was suggested that $BC_5$-III where the isolated boron atoms are uniformly distributed is the most stable boron doped graphene structure.[18,19] However, our calculations show that $BC_5$-III is energetically less favorable and has a much higher energy than $BC_5$-I by 51meV/atom. This might be due to the fact that $BC_5$-I has more C-C π bonds than $BC_5$-III, similar to the cases of the hydrogenation and oxidation of graphene, where hydrogen atoms or oxygen atoms prefer to staying close to each other in graphene.[32,33] However, we find that the full phase separation does not occur in boron doped graphene structures because the 2D graphene-like boron structure has a higher energy. Different from $BC_2$ and $BC_5$, $BC_3$ with 1D zigzag boron chains is not the most stable 2D structure. Instead, the structure with uniformly distributed boron atoms ($BC_3$-I) is more favorable with a lower energy of

52meV/atom. Below we will show that the stability of $BC_3$-I is originated from its peculiar semiconducting electronic structure. $BC_3$-I with two boron atoms at 1 and 4 positions of the carbon six-membered ring was previously observed experimentally.[15] For the first time, we confirm by global optimization technique that $BC_3$-I is indeed the most stable 2D structure.

For BC, the stable 2D structure (BC-I in Fig.4) is strip-like with alternative boron chains and armchair carbon chains. Every carbon atom is $sp^2$ hybridized, forming two C-C bonds and one B-C bond. Each boron atom has four neighbors with one B-C bond and three B-B bonds. The boron atoms in BC-I can be viewed as connected prismatic $B_4$. Carbon atoms in the metastable BC structures (BC-II and BC-III in Fig.4) are either three-fold or four-fold bonded.

The lowest energy 2D structure for $B_2C$ can be seen as the addition of boron atoms to the center of the $B_4C_2$ six-membered ring (with C at the opposite side of the ring) of the $B_5C_3$ graphene structure. As a result, boron atoms form six-, four-, and three-fold bonds. Our calculations also find the $B_2C$ structure with the ptC moiety ($D_{2h}$ symmetry, $T_{11}$ in the notation by Pei and Zeng)[34] previously proposed by Wu et al..[20] However, this structure has a higher energy (49meV/atom) than the most stable structure ($B_2C$-I). It is interesting to note that the most stable $B_2C$ structure contains the ptC atoms but with a different motif (approximate $C_{2v}$ symmetry, see Fig. 11b). This intriguing ptC motif was earlier proposed to be present in a metastable configuration of $C_3B_8$ ($T_7$ in the notation by Pei and Zeng).[34] Here, we show that the $T_7$ ptC motifs can in fact be stabilized in a stable 2D structure of $B_2C$. Previously, experiments have indicated that some aluminum-containing ptC species in the gas phase, such as $CAl_4^-$, $CAl_3Si^-$, $CAl_3Ge^-$, neutral $CAl_3Si$, and neutral $CAl_3Ge$, have been isolated and detected in photoelectron spectroscopy experiments.[35] The boron-containing ptC species have not been observed. This might be due to the fact that most experimental work focused on the carbon-rich 2D B-C compounds. Future experiments on boron-rich 2D B-C systems might confirm our prediction of this novel ptC motif.

The most stable 2D $B_3C$ structure ($B_3C$-I) is shown in Fig. 6. This structure contains alternative zigzag boron chains and zigzag boron-carbon chains. And there are similar ptC motifs as that in $B_2C$-I. It is noteworthy that the two metastable 2D $B_3C$ structures ($B_3C$-II and $B_3C$-III) are closely related to $B_3C$-I: $B_3C$-II can be obtained by moving half of the boron atoms in the zigzag boron-carbon chain toward the zigzag boron chain to form more B triangles; The zigzag boron-carbon chains in $B_3C$-III are equivalent as a result of swapping a carbon atom in the zigzag boron-carbon chain with its neighboring boron atom in the zigzag boron chain in $B_3C$-I. The most stable 2D structure of $B_5C$ (Fig. 7) is similar to $B_3C$-I except that the ribbon width of the triangular boron sheet is now four instead of two. In the metastable structures ($B_5C$-II and $B_5C$-III), all carbon atoms are three-fold coordinated. Interestingly, there is a large hole formed by six boron atoms and two carbon atoms in $B_5C$-III. To summarize, boron-rich 2D B-C compounds have peculiar ptC motifs and boron triangles. And they do not resemble clearly the structural feature of α-sheet boron structures, in contrast to the carbon-rich case.

In our structural searches, we constrain the systems to be an exact 2D monolayer structure. It might be possible that the predicted structures are not stable against out-of-plane distortions. We therefore perform phonon calculations to check the dynamic stability of the predicted stable 2D structures. Our calculations show that all the 2D structures have no appreciable unstable phonon modes except for the 2D $B_3C$-I and $B_5C$-I structures. $B_3C$-I has a large imaginary frequency (6.7i THz) at zone center Γ point. By distortion of the atoms along the vibrational eigenvectors of the zone center soft phonon mode, we derive a stable structure with a 0.42 Å buckling which is more stable than the exact 2D $B_3C$-I structure by 24 meV/atom. Due to the structural similarity, a similar distortion occurs in $B_5C$-I.

We also perform first-principles molecular dynamics (MD) simulations to examine the thermal stability of the 2D structures. The constant temperature and volume (NVT) ensemble was adopted. The time step is 3fs and the total simulation time is 15ps for each given temperature. We find that almost all the lowest energy 2D structures are stable up to 1000K. In particular, the structure of BC-I remains almost intact at 2000 K [see Fig. 8b]. The high thermal stability of BC-I should be due to the high stability of carbon armchair chain and the fact that each boron has one relatively strong B-C bond. However, $B_3C$-I is unstable at 1000 K: Some B-C bonds in the zigzag B-C chains might be broken to form more B-B bonds, resulting in a motif similar to that in $B_3C$-II [Fig. 8a]. This is not unreasonable since $B_3C$-II is structurally closely related to $B_3C$-I and has only a slightly higher energy than $B_3C$-I, with which only a small kinetic barrier between $B_3C$-I and $B_3C$-II is expected.

Our electronic band structure calculations show that all the 2D B-C compounds are metallic except for $BC_3$-I (See Fig.9 for representative band structures). The metallicity is stemmed by the delocalized $2p_z$ π electrons of carbon and boron atoms, similar to those in graphene and boron 2D α-sheet.[11] The LDA calculation shows that $BC_3$-I has an indirect gap of 0.52eV with the valence band maxima at Γ, which is in agreement with the reported value (0.66eV) from a local orbital pseudopotential calculation.[17] It is interesting to note that we recently found 2D $NC_3$ has the same structure as $BC_3$-I and is also a semiconductor.[36] The semiconducting behavior is due to six-membered "benzene" rings isolated by boron atoms: The boron atom is $sp^2$ hybridized and all its three 2p electrons are participated in the formation of σ-bonds; For the "benzene" ring, there are three bonding orbitals and three antibonding orbitals separated via an energy gap.

The partial density of states (PDOS) of the predicted 2D B-C compounds is analyzed carefully. By the 2D nature, there is a mirror plane symmetry (the basal plane) in each of the 2D systems. The $p_z$ orbital is odd with respect to the basal plane, while s, $p_x$, $p_y$ orbitals are even. Therefore, there is no mixing between out-of-plane $p_z$ π states and in-plane $s+p_x+p_y$ σ states. The representative PDOSs for $BC_2$-I, BC-I, and $B_2C$-I are shown in Fig. 10. As mentioned above, the $p_z$ π states are partially occupied. We can clearly see that the separation between bonding σ states and anitibonding σ* states. For BC$_2$-I [Fig. 10(a)], the C σ states end at -1.7 eV and the C σ* states start at 7.0 eV; while the B σ states extend up to 6 eV. For BC$_2$-I [Fig. 10(c)], the hybridized C-B σ states are hole-doped with the Fermi level 0.8 eV below the top of the σ bonding states. Interestingly, in BC-I, the σ bonding states are fully occupied, while the anitibonding σ* states are empty [see Fig. 10(a)]. The large energy separation (4.2 eV) between σ and σ* states and the full occupation of the σ bonding states are responsible for the peculiar stability of BC-I. Previously, it was found[11] that boron α-sheet is most stable due to the optimal filling of σ bonding states: Electrons fill all σ bonding states while leaving all antibonding σ* states empty, and any remaining electrons partially fill out-of-plane π states. The reason why best boron 2D structures are determined by the optimal filling of the in-plane manifold is that σ bonds are stronger than π bonds. In this study, we find that the most stable 2D structures of BC and BC$_3$ (PDOS for the semiconducting BC$_3$-I not shown here) have such an optimal filling of σ bonding states, while the other 2D B-C compounds do not.

To gain more insight into the chemical bonding in 2D boron-carbon nanostructures, we plot the electron localization function (ELF) for representative systems using the formulation of Silvi and Savin.[37] The topological analysis of ELFs can be used to classify chemical bonds rigorously.[38] Because the σ states are more localized than π states, the ELF distribution with a large value (e.g., 0.7) for B-C compounds mostly reflects the in-plane σ states. The ELF isosurface (0.7) plot [Fig. 11a] for BC$_2$-I shows that the p orbitals of the 1D zigzag boron chain are rather delocalized, which might explain the stability of the 1D zigzag boron chain motif [lower panel of Fig. 11(a)]. For BC-I, there are four C-C σ bonds, four B-C σ bonds, one B-B two center σ bond, and two three-center B bonds in each unit cell (each unit cell has four B atoms and four C atoms). The total number (11) of σ bonds is consistent with the number (22) of occupied σ electrons integrated from the in-plane PDOS. The other 6 electrons occupy the π states.

For B$_2$C-I, ELF is significantly distributed around the center of the boron triangles, indicating the presence of the boron three-center bond. Around each three-fold coordinated carbon atom, the ELF localizes at the bond centers of the three B-C bonds, similar to the ELF distribution of graphene (see Fig. 11a for the ELF of the graphene part of BC$_2$). For the ptC atom in B$_2$C-I, the ELF mainly distributes between carbon and B$_1$, B$_2$, and B$_3$ (lower panel of Fig. 11c), whereas the bonding between the ptC atom and B$_4$ is weak. For comparison, we also show the ELF for B$_2$C-IV with D$_{2h}$ ptC motifs (lower panel of Fig. 11d). In this case, the ELF is symmetrically distributed along each B-C bond. We note that there is no boron three-center bond in B$_2$C-IV, which might explain the physical origin of its higher energy. It was recently shown that there is no two-center σ B-B bond in the α-sheet boron structure.[39] Here, we find that there are two-center σ B-B bonds in 2D B-C compounds, especially in some boron-rich B-C compounds (see Figs. 11b and 11c).

## Conclusion

To predict 2D nanostructures, we develop a global optimization method based on the particle-swarm optimization (PSO) algorithm as implemented in CALYPSO code. Using the PSO algorithm, we predict new stable structures of 2D B-C compounds for a wide range of boron concentrations. For some of the system such as BC$_5$ and B$_2$C, the predicted structures have much lower energies than previous proposed structures. Our calculations show that: (1) almost all 2D B-C compounds are metallic except for BC$_3$ which is a magic case where the isolation of carbon six-membered ring by boron atoms results in a semiconducting behavior; (2) for C-rich B-C compounds, the most stable 2D structures can be viewed as boron doped graphene structures. Usually boron forms 1D zigzag chain except for BC$_3$ where boron atoms are uniformly distributed; (3) the most stable 2D structure of BC has alternative carbon and boron ribbons with strong in-between B-C bonds, resulting in a high thermal stability above 2000 K; (4) for B-rich 2D B-C compounds, there is a novel planar-tetracoordinate carbon motif with an approximate C$_{2v}$ symmetry. ELF analysis of the nature of the bonding shows that the delocalizd p states of 1D zigzag boron chain in the carbon-rich compounds and the three-center boron bonds in the boron-rich case play an important role in the structural stability. The new 2D B-C compounds predicted in this work might be promising candidates for nanoelectronics applications, energy materials (electrode in Li-ion battery, hydrogen storage, and cheap catalysis in fuel cell). From the predicted 2D B-C compounds, one can derive many new nanostructures (e.g., nanoribbons, nanotubes, nanocages) which might have a wealth of exotic electronic and/or magnetic properties.

## Acknowledgment

Work at Fudan was partially supported by NSFC, the Research Program of Shanghai Municipality and the Special Funds for Major State Basic Research, Pujiang plan, and Program for Professor of Special Appointment (Eastern Scholar). Work at Jilin University was supported by NSFC under No. 91022029 . Work at NREL was supported by U.S. DOE under Contract No. DE-AC36-08GO28308.

*Electronic address: hxiang@fudan.edu.cn

**Supporting Information Available:** Complete ref 5.

## References

[1] Novoselov, K. S.; Geim, A. K.; Morozov, S. V.; Jiang, D.; Zhang, Y.; Dubonos, S. V.; Grigorieva, I. V.; Firsov, A. A. *Science.* **2004**, *306*, 666.
[2] Novoselov, K. S.; Geim, A. K.; Morozov, S. V.; Jiang, D.; Katsnelson, M. I.; Grigorieva, I. V.; Dubonos, S. V.; Firsov, A. A. *Nature.* **2005**, *438*, 197.
[3] Novoselov, K. S.; McCann, E.; Morozov, S. V.; Falko, V. I.; Katsnelson, M. I.; Zeitler, U.; Jiang, D.; Schedin, F.; Geim, A. K. *Nat Phys.* **2006**, *2*, 177.
[4] Zhang, Y. B.; Tan, Y. W.; Stormer, H. L.; Kim, P. *Nature.* **2005**, *438*, 201.
[5] Coleman, J. N. *et al. Science.* **2011**, *331*, 568.
[6] Kunstmann, J.; Quandt, A. *Phys. Rev. B* **2006**, *74*, 035413.
[7] Lau, K. C.; Pandey, R.; Pati, R.; Karna, S. P. *Appl. Phys. Lett.* **2006**, *88*, 212111.

[8] Ciuparu, D.; Klie, R. F.; Zhu, Y.; Pfefferle, L. *J. Phys. Chem. B* **2004**, *108*, 3967.
[9] Boustani, I.; Quandt, A. *Europhys. Lett.* **1997**, *39*, 527.
[10] Kunstmann, J.; Quandt, A.; Boustani, I. *Nanotechnology*. **2007**, *18*, 155703.
[11] (a) Tang, H.; Ismail-Beigi, S. *Phys. Rev. Lett.* **2007**, *99*, 115501. (b) Tang, H.; Ismail-Beigi, S. *Phys. Rev. B* **2009**, *80*, 134113.
[12] Yang, X.; Ding, Y.; Ni, J. *Phys. Rev. B* **2008**, *77*, 041402(R).
[13] Szwacki, N. G.; Sadrzadeh, A.; Yakobson, B. I. *Phys. Rev. Lett.* **2007**, *98*, 166804.
[14] Saxena, S.; Tyson, T. A. *Phys. Rev. Lett.* **2010**, *104*, 245502.
[15] Kouvetakis, J.; Kaner, R. B.; Sattler, M. L.; Bartlett, N. J. *Chem. Soc. Chem. Commun.* **1986**, 1758.
[16] Yanagisawa, H.; Tanaka, T.; Ishida, Y.; Matsue, M.; Rokuta, E.; Otani, S.; Oshima, C. *Phys. Rev. Lett.* **2004**, *93*, 177003.
[17] Tománek, D.; Wentzcovitch, R. M.; Louie, S. G.; Cohen, M. L. *Phys. Rev. B* **1988**, *37*, 3134.
[18] Way, B. M.; Dahn, J. R.; Tiedje, T.; Myrtle, K.; Kasrai, M. *Phys. Rev. B* **1992**, *46*, 1697.
[19] Hu, Q.; Wu, Q.; Ma, Y.; Zhang, L.; Liu, Z.; He, J.; Sun, H.; Wang, H. -T.; Tian, Y. *Phys. Rev. B* **2006**, *73*, 214116.
[20] Wu, X.; Pei, Y.; Zeng, X. C. *Nano Lett.* **2009**, *9*, 1577.
[21] (a) Wang, Y.; Lv, J.; Zhu, L.; Ma, Y. *Phys. Rev. B* **2010**, *82*, 094116. (b) Http://nlshm-lab.jlu.edu.cn/~calypso.html.
[22] (a) Oganov, A.; Glass, C. *J. Chem. Phys.* **2006**, *124*, 244704. (b) Trimarchi, G.; Zunger, A. *Phys. Rev. B* **2007**, *75*, 104113. (c) Pickard, C.; Needs, R. *Phys. Rev. Lett.* **2006**, *97*, 045504. (d) Deaven, D. M.; Ho, K. M. *Phys. Rev. Lett.* **1995**, *75*, 288. (e) Xiang, H. J.; Wei, S. –H.; Gong, X. G. *J. Am. Chem. Soc.* **2010**, *132*, 7355. (f) Call, S. T.; Zubarev, D.; Yu. Boldyrev, A. I. *J. Comput. Chem.* **2007**, *28*, 1177.
[23] Kennedy, J.; Eberhart, R. *Proceedings of IEEE Inter. Conf. on Neural Networks. IV*. **1995**, 1942.
[24] Lv, J.; Wang, Y.; Zhu, L.; Ma, Y. *Phys. Rev. Lett.* **2011**, *106*, 015503.
[25] Li, P.; Gao, G.; Wang, Y.; Ma, Y. *J. Phys. Chem. C* **2010**, *114*, 21745.
[26] Zhu, L.; Wang, H.; Wang, Y.; Lv, J.; Ma, Y. *Phys. Rev. Lett.* **2011**, *106*, 145501.

[27] (a) Blöchl, P. E. *Phys. Rev. B* **1994**, *50*, 17953. (b) Kresse, G.; Joubert, D. *ibid*, **1999**, *59*, 1758.

[28] Kresse, G.; Furthmüller, J. *Comput. Mater. Sci.* **1996**, *6*, 15; *Phys. Rev. B* **1996**, *54*, 11169.

[29] Togo, A.; Oba, F.; Tanaka, I. *Phys. Rev. B* **2008**, *78*, 134106.

[30] Song, L.; Ci, L. J.; Lu, H.; Sorokin, P. B.; Jin, C. H.; Ni, J.; Kvashnin, A. G.; Kvashnin, D. G.; Lou, J.; Yakobson, B. I.; Ajayan, P. M. *Nano Lett.* **2010**, *10*, 3209.

[31] Tusche, C.; Meyerheim, H. L.; Kirschner, J. *Phys. Rev. Lett.* **2007**, *99*, 026102.

[32] Xiang, H. J.; Kan, E. J.; Wei, S.-H.; Whangbo, M.-H.; Yang, J. L. *Nano Lett.* **2009**, *9*, 4025.

[33] Xiang, H. J.; Wei, S.-H.; Gong X. G. *Phys. Rev. B* **2010**, *82*, 035416.

[34] Pei, Y.; Zeng, X. C. *J. Am. Chem. Soc.* **2008**, *130*, 2580.

[35] (a) Li, X.; Wang, L.-S.; Boldyrev, A. I.; Simons, J. *J. Am. Chem. Soc.* **1999**, *121*, 6033. (b) Wang, L.-S.; Boldyrev, A. I.; Li, X.; Simons, J. *J. Am. Chem. Soc.* **2000**, *122*, 7681.

[35] Xiang, H. J.; Huang, B.; Li, Z.; Wei, S. –H.; Yang, J.; Gong, X. G., arXiv:1105.3540v1 (2011).

[36] Silvi, B.; Savin, A. *Nature.* **1994**, *371*, 683.

[37] See for example, Li, Z.; Yang, J.; Hou, J. G.; Zhu, Q. *Angew. Chem., Int. Ed.* **2004**, *43*, 6479.

[38] Galeev, Timur R.; Chen, Q.; Guo, J.-C.; Bai, H.; Miao, C.-Q.;Lu, H.-G.; Sergeeva, A. P.; Li, S.-D.; Boldyrev, A. I. *Phys. Chem. Chem. Phys.* **2011**, *13*, 11575.

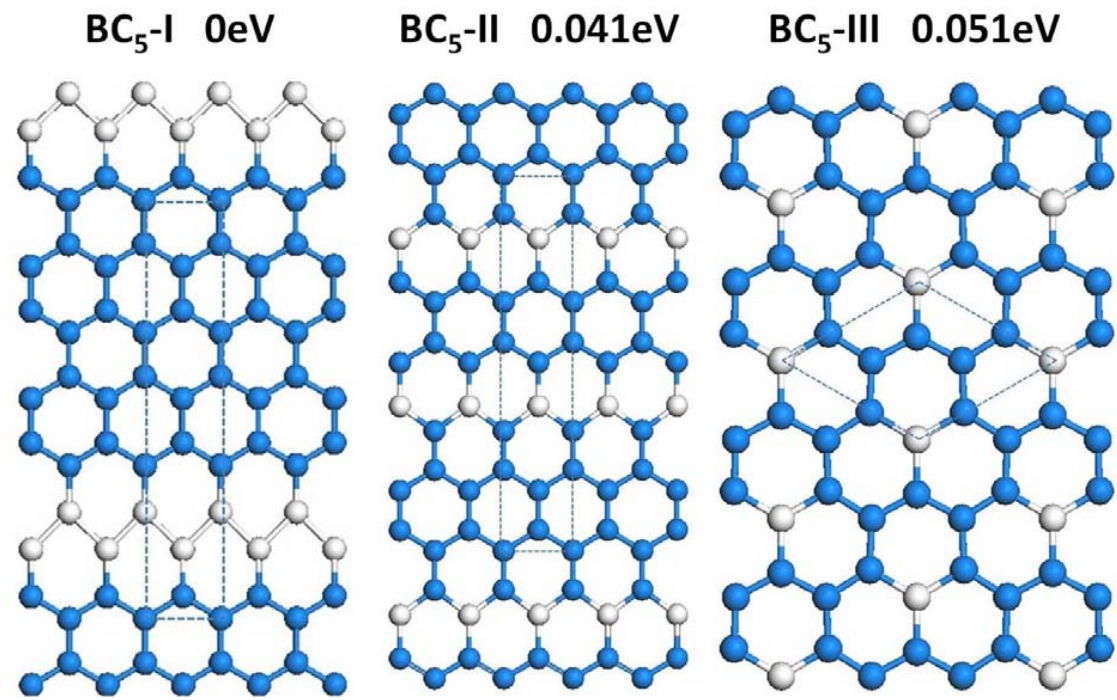

**FIG. 1:** Low energy 2D structures of $BC_5$ from the PSO simulations. The blue (dark) atom is C and the gray (light) atom is B. The relative energy per atom is indicated.

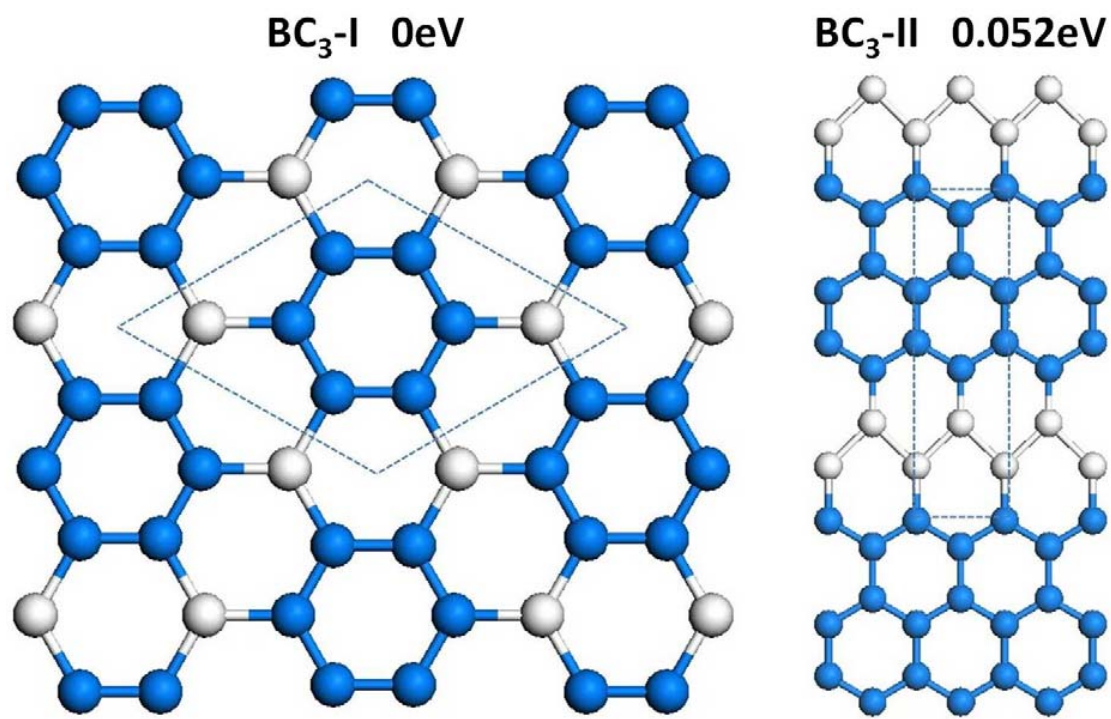

**FIG. 2:** Low energy 2D structures of $BC_3$ from the PSO simulations.

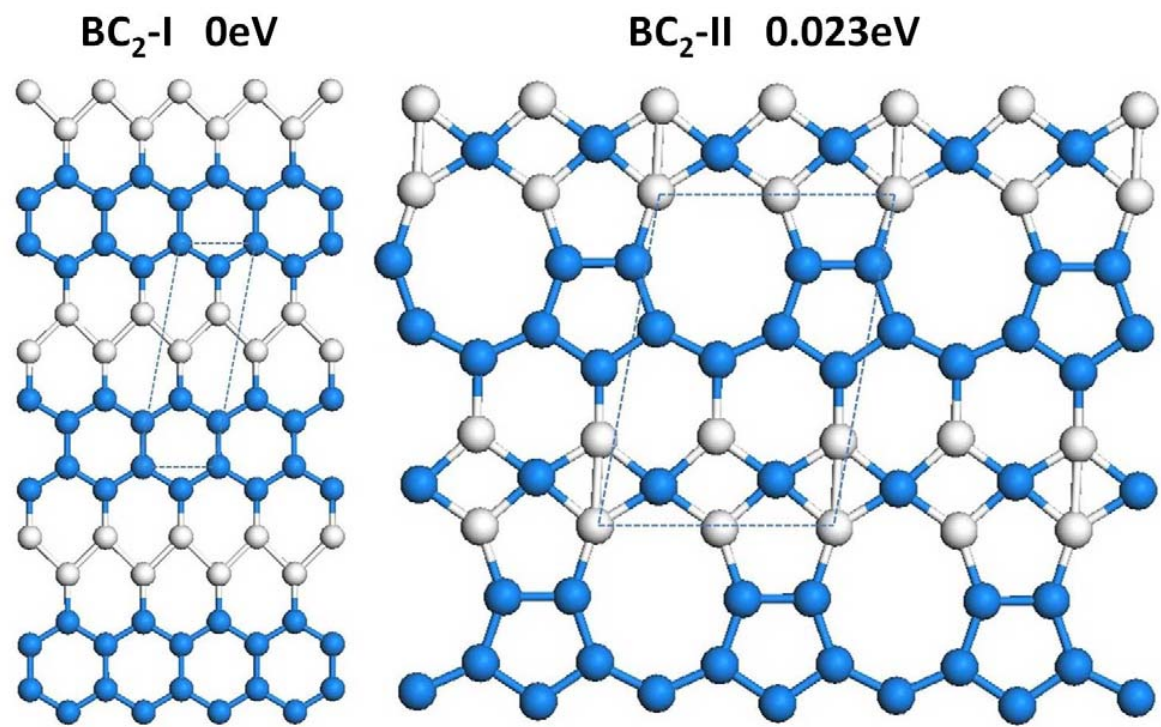

FIG. 3: Low energy 2D structures of $BC_2$ from the PSO simulations.

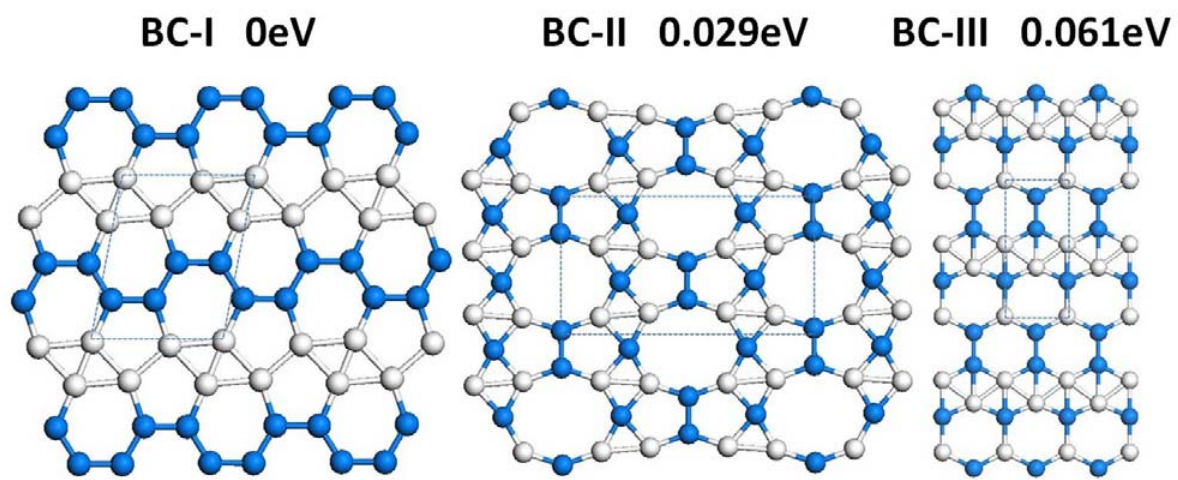

FIG. 4: Low energy 2D structures of BC from the PSO simulations.

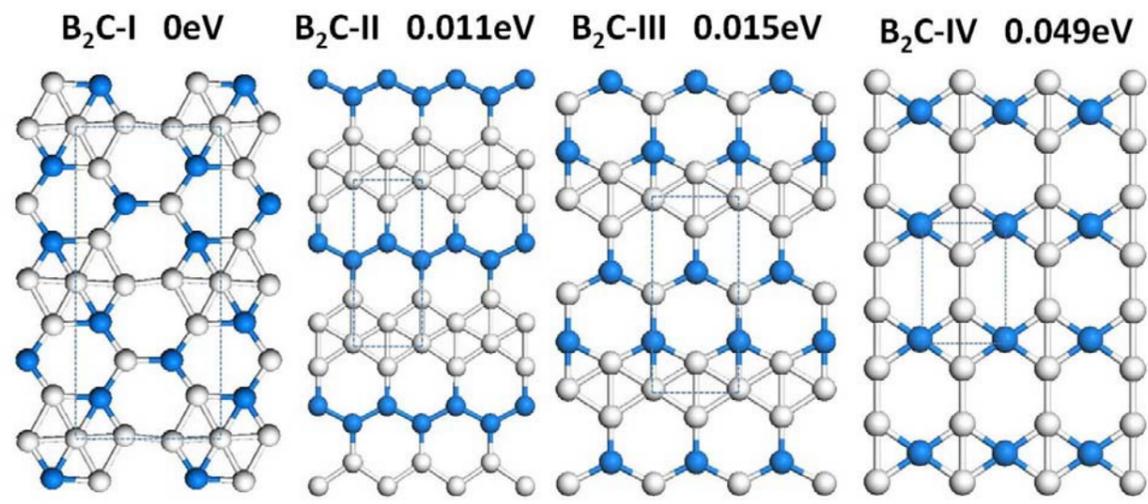

**FIG. 5:** Low energy 2D structures of B$_2$C from the PSO simulations.

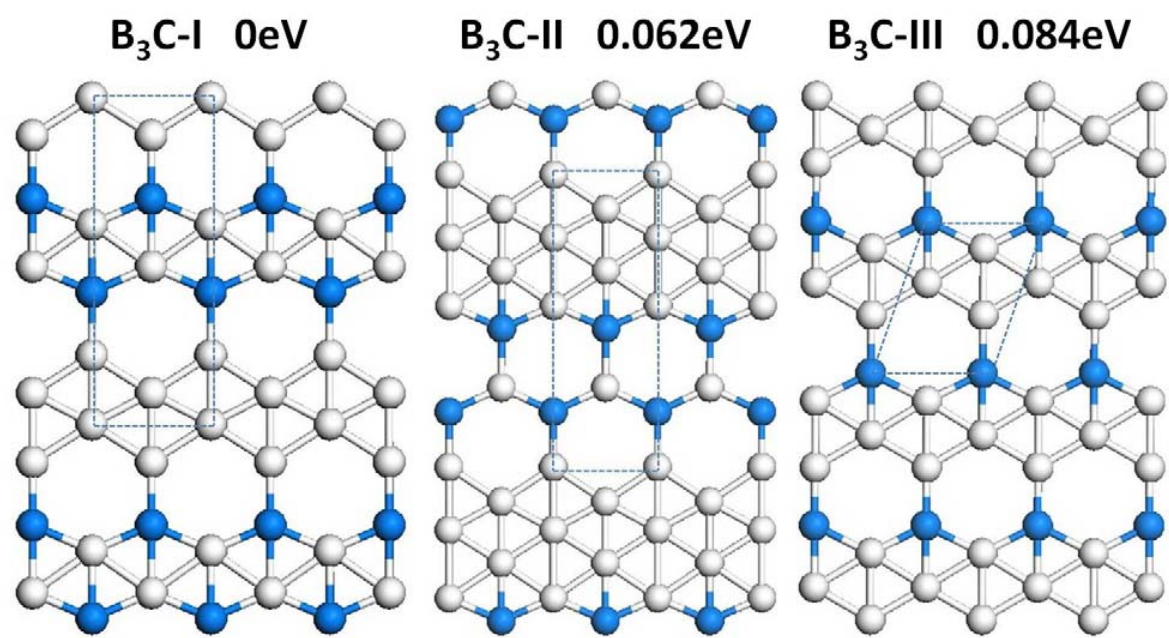

**FIG. 6:** Low energy 2D structures of B$_3$C from the PSO simulations.

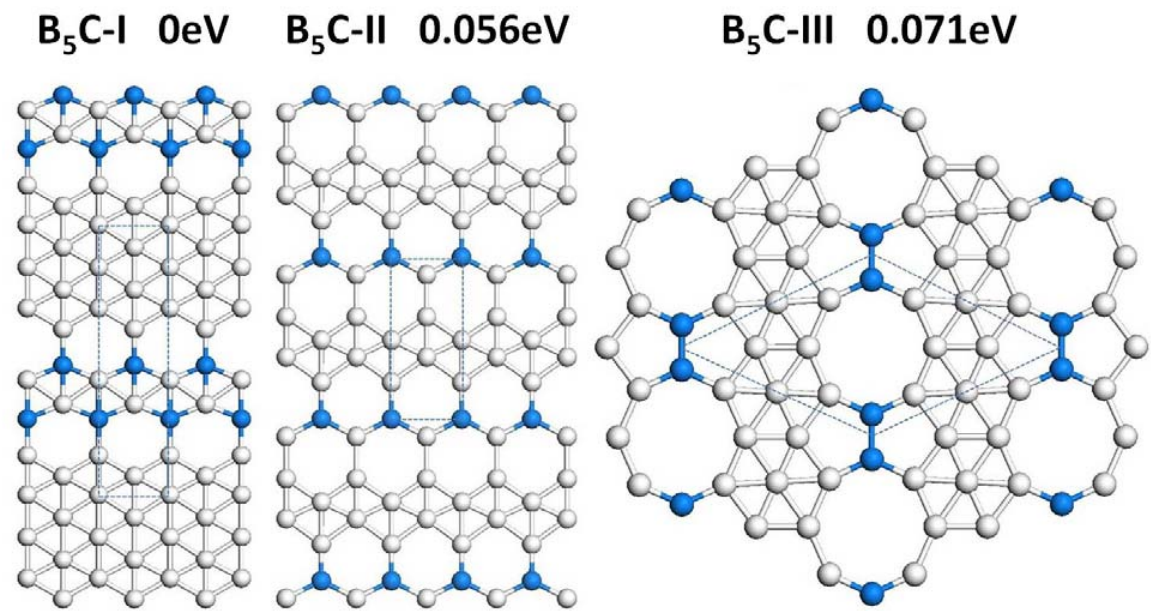

FIG. 7: Low energy 2D structures of $B_5C$ from the PSO simulations.

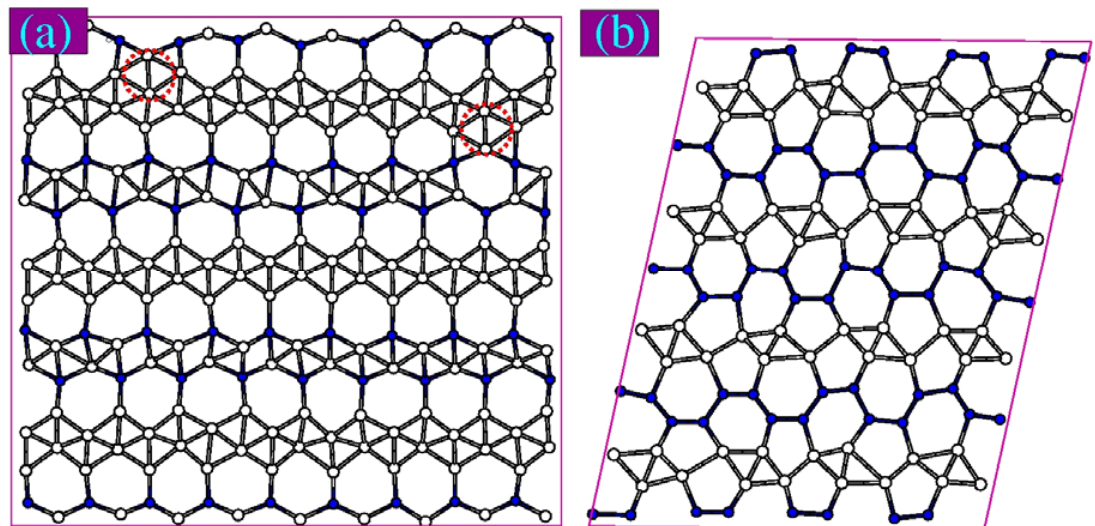

FIG. 8: (a) Snapshot of $B_3C$-I after a 15 ps MD simulation at 1000 K. The dashed circles denote the newly formed B-B bonds. (b) Snapshot of BC-I after a 15 ps MD simulation at 2000 K.

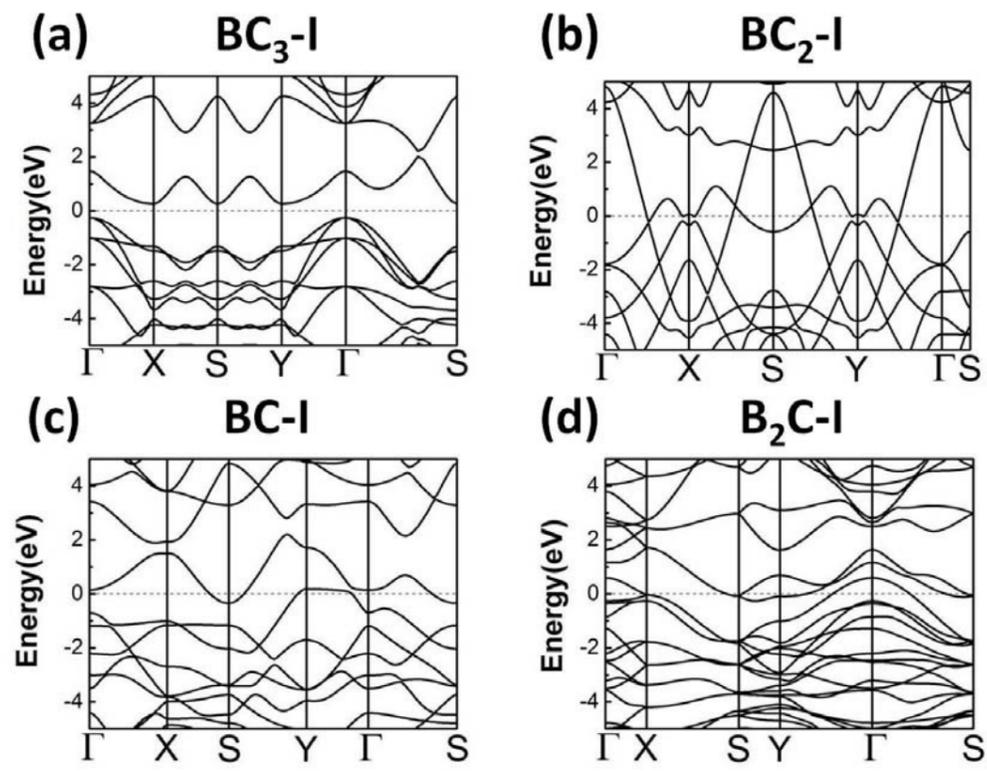

**FIG.9:** LDA band structures of (a) $BC_3$-I, (b) $BC_2$-I, (c) BC-I, and (d) $B_2C$-I.

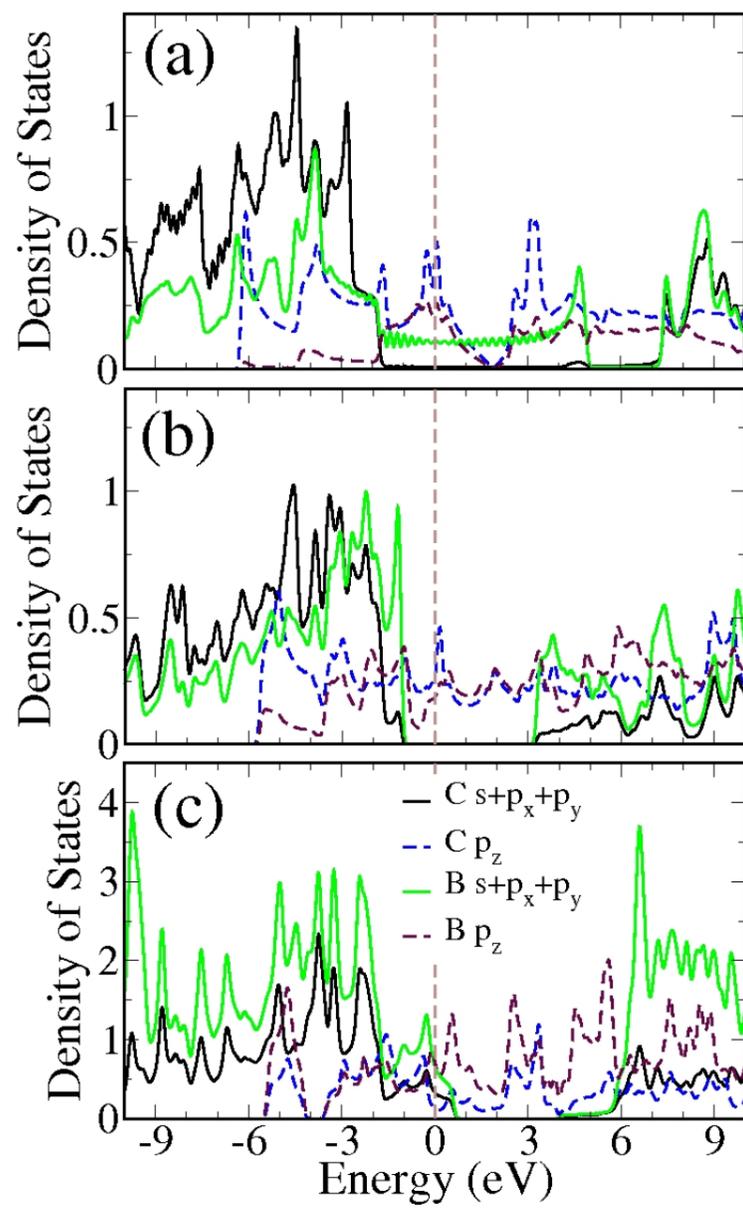

**FIG. 10: Partial density of states for (a) BC$_2$-I, (b) BC-I, and (c) B$_2$C-I. The vertical dashed lines denote the Fermi level.**

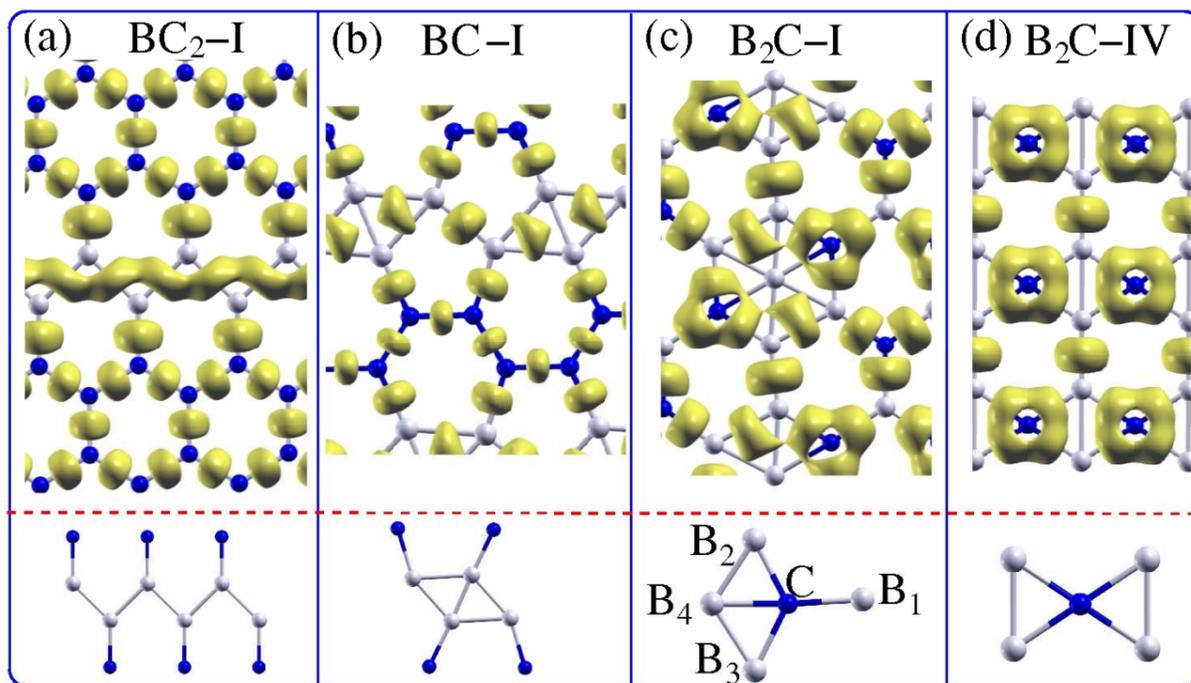

**FIG.11: Isosurfaces of electron localization function with the value of 0.7 for (a) BC$_2$-I, (b) B$_2$C-I, and (c) B$_2$C-IV. The structural characters are shown in the lower panels.**

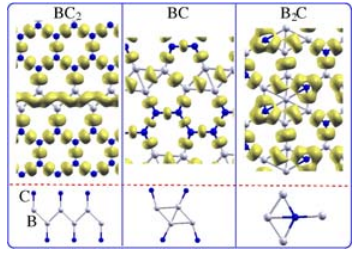

**Toc graphic**